\documentclass[twocolumn,showpacs]{revtex4}
\usepackage[pctex32]{graphicx}

\begin{document}

\title{EXPERIMENTAL AND COMPUTATIONAL STUDY OF THE EFFECT OF THE 
SYSTEM SIZE ON ROUGH SURFACES FORMED BY SEDIMENTING PARTICLES IN 
QUASI-TWO-DIMENSIONS}

\author{K. V. McCloud}
\affiliation{Department of Physics and Engineering,
Xavier University of Louisiana, New Orleans, LA 70125}

\author{U. Cardak}
\author{M. L. Kurnaz}
\affiliation{Department of Physics, Bogazici University, 34342
Bebek Istanbul}

\begin{abstract}

The roughness exponent of surfaces obtained by dispersing silica 
spheres into a quasi-two-dimensional cell is examined using 
experimental and computational methods. The cell consists of two 
glass plates separated by a gap, which is comparable in size to the 
diameter of the beads. We have studied the effect of changing the 
gap between the plates to a limit of about twice the diameter of 
the beads. If the conventional scaling analysis is performed, the 
roughness exponent is found to be robust against changes in the gap 
between the plates. The surfaces formed have two roughness exponents 
in two length scales, which have a crossover length about 1 cm.; 
however, the computational results do not show the same crossover 
behavior. The single exponent obtained from the simulations stays 
between the two roughness exponents obtained in the experiments. 
\end{abstract}

\maketitle

\section{INTRODUCTION\protect\\ }
\label{sec:level1}

The formation of rough surfaces through the sedimentation of particles 
through a viscous fluid is a complex problem, but one with many 
applications, ranging from the study of fundamental non-equilibrium 
statistical physics to various industrial processes such as the growth 
of films by deposition \cite{Vicsek7900, Barabasi5260}.   The 
presence of the viscous fluid allows both particle/particle interactions 
as well as particle/wall interactions during sedimentation which are 
not normally considered in deposition processes, but which may well 
be present in actual systems of interest and which it can be assumed 
will have an effect on the final surface. Surfaces formed by sedimentation 
are close to the original problem of sedimentation of particles sedimenting 
along straight vertical trajectories first studied by Edwards and 
Wilkinson \cite{Edwards11280}. However, the hydrodynamic particle/particle 
and particle/wall forces are in principle long-range, making the rough 
surfaces formed by particles sedimenting in a viscous fluid are a different 
growth situation from the simpler vertical deposition. The situation is 
further complicated by the presence of back flows of fluid caused by the 
motion of significant numbers of particles \cite{Davis10770, Auzerais9790, 
Schwarzer5600}.

In this work we are primarily interested studying the effect 
of the particle-wall interactions on the roughness of the final 
interface. The problem of the motion of a sphere parallel to a 
single wall as the limiting case of motion of a small sphere in a 
cylindrical container when the sphere approaches the cylinder wall 
and the more general one of the motion of a sphere parallel to two 
external walls were treated by Faxen \cite{Faxen12520}. 
Unfortunately, the exact nature of the interactions between the 
particles in the presence of the walls is very difficult to determine. 
Analytical sedimentation theory has succeeded in analyzing the 
effective settling velocity \cite{Smoluchowski12540, Smoluchowski12550,
Burgers12430, Brenner12370, Kynch12360, Hasimoto12350, Faxen12510}
and velocity fluctuations \cite{Difelice21410, Ladd14220} of 
particles in a dilute regime in the presence of the walls  and some 
features of many-body interactions between the particles 
\cite{Vansaarlos11200, Mazur11310} when there are no walls 
\cite{Happel12270, Tee1670, Mucha21160, Mucha21330}. Recent 
theoretical \cite{Auzerais9790, Brady9830, Koch8420, 
Cichocki13780, Ladd14080, Jones14070, Kuusela13760, Dance21170,
Dance21190, Miguel1990, Schwarzer21360, Xu21200} and experimental 
\cite{Nicolai5490, Xue7950, Segre4660, Peysson21390, Peysson21340,
BernardMichel21250, Dufresne21310, Guazzelli21280, Herzhaft3810, 
Rouyer21320, Segre1590} work hold out some hope of determining 
effective particle interactions through a wide range of volume 
fractions and Peclet numbers in sedimentation problems. Also 
simulations of deposition of elongated particles \cite{Asikainen13870, 
Hellen21540, Provatas13770, Provatas13970, Provatas13980, 
Vinnurva13930} indicate that application of sedimentation problems 
is not restricted to spherical particles, but may well expand into 
areas like the paper industry. New developments in the effect of the 
container size on the divergence of the velocity fluctuations 
\cite{Brenner21370, Climent21210, Cunha14020, Cunha21230, Ladd14060, 
Levine21380} show the importance of the presence of the wall in 
determining the particle-particle and particle-wall interactions.

In our previous work \cite{Kurnaz6550, Kurnaz13030, Mccloud12950, 
Mccloud12910} on the quasi-one-dimensional surfaces formed by 
particles sedimenting through a viscous fluid in a quasi-two-dimensional 
cell we have found that the surfaces formed by sedimenting particles 
are rough on all length scales between the particle size and the cell 
size. Using the scaling ansatz proposed by Family and Vicsek 
\cite{Family10790} discussed below, it was found that different 
roughness exponents were found in two different length-scale regimes, 
with a crossover length scale. These roughness exponents and the 
crossover length scale have been found to be independent of the cell 
aspect ratio or the viscosity of the fluid through which the particles 
settle \cite{Kurnaz13030}. The exponent found at long length scales 
has been shown to depend on the rate at which particles are deposited 
into the cell (hence to the strength of the interaction between the 
particles) \cite{Mccloud12950}. This lead to the conclusion that the 
scaling exponent seen at long length scales depended on the details 
of the hydrodynamic interactions between the particles, while the 
exponent seen at small length scales, which remained relatively 
unaffected by changes in the deposition rate, may be due to more 
universal considerations.

In this work, we have investigated the effect of the particle-wall 
interactions on the roughness of the final interfaces formed by 
quasi-two dimensional sedimentation of small glass beads through a 
viscous fluid. Simulations of the same system are compared to experimental 
results with the aim of untangling the effects of the viscous fluid on 
the process from the better understood effects of the deposition process.

\section{EXPERIMENTAL WORK\protect\\ }
\label{sec:level2} 

\begin{figure}[!]
\includegraphics[width=8cm]{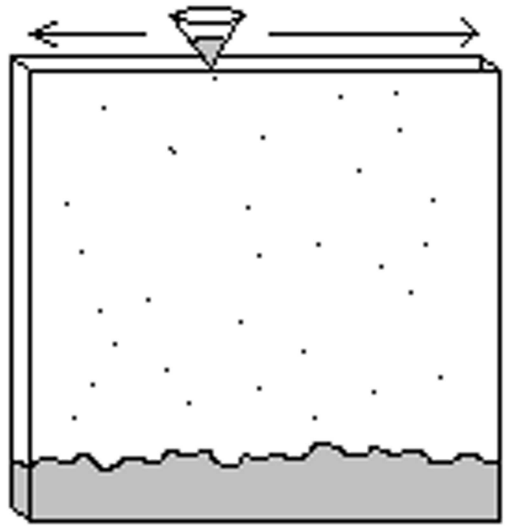}
\caption{The sedimentation cell consists of two glass plates with
a small gap between them, of the order of the diameter of the
glass beads.   The cell is filled with a viscous fluid such as
oil, and a funnel sweeps across the top of the cell, delivering a
mixture of oil and beads to the cell.   The beads settle to the
bottom of the cell and build a rough surface.} \label{fig1}
\end{figure}

In our previous work \cite{Kurnaz6550, Kurnaz13030, Mccloud12950, 
Mccloud12910}, sedimentation experiments in quasi-two dimensions have 
been carried out using two different types of cells, denoted as "closed" 
and "open" cells. Closed cells were constructed of 1/4 in. float glass, 
held 1 mm apart by sealed side frames of precision machined Plexiglas. 
Around 10,000 0.6 mm-diameter monodisperse silica spheres were placed in the 
cell, which was then filled with a viscous fluid (such as glycerin or 
paraffin oil) and closed. Each cell could be rotated about a horizontal 
axis perpendicular to the gap direction. When the cell was rotated, the 
particles which had been at rest on the bottom fell though the viscous fluid, 
slowly building up a new surface at the bottom of the cell. In the closed 
cells we only had a fixed gap size between the cell walls, but we had 
different sizes of cells filled with fluids of different viscosity. However 
the number of the particles sedimenting at any time could not be controlled. 
The open cell (which was the one used in this experimental work) was also 
constructed of 1/4 in. float glass, separated by strips of Teflon of known 
thickness. It had dimensions comparable to the closed cell, but was open 
at the top, so that beads could be dispersed through a funnel which steadily 
dropped beads as it traveled back and forth across the top of the cell 
(Fig.~\ref{fig1}). In this way, the deposition rate of the beads into the cell 
could be controlled precisely by varying the speed and the size of the 
funnel. The cell could be taken apart and a different thickness of Teflon 
inserted to change the gap between the plates.   

\begin{figure}[!]
\includegraphics[width=8cm]{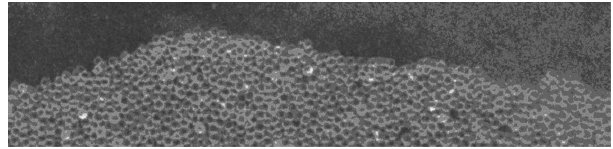}
\caption{Portion of a rough surface formed by 0.6 mm diameter
glass beads sedimenting through heavy paraffin oil.} \label{fig2}
\end{figure}

We have tested the effect of changing the distance between the walls on the 
roughness of the final interface while keeping the particle density fairly 
constant. The experiments took place in the open cell and we investigated 
the effect of variability in the gap by setting the gap at different values, 
and measuring the effect of the walls on the roughness of the final interface 
formed during sedimentation. We investigated gaps ranging from 
0.8 mm to 2.0 mm, while the particle size was kept constant at 0.6 mm. 
The ratio of the gap thickness to the bead diameter was defined as a 
dimensionless parameter $R$, and our experiments spanned a gap/bead diameter 
ratio of $R = 1.33$ to $R = 3.33$. In all cases, the deposition rate of beads 
into the cell was controlled at about 4 beads/sec so the average distance 
between the particles was about $20R$. The surface was photographed during 
and at the end of the deposition process and the photographs were digitized 
by a Nikon LS-2000 film scanner. Individual particles were typically resolvable 
and thus the position of the particles on the interface could be traced 
accurately (Fig.~\ref{fig2}). There is a limit to the extent over which the 
gap can be widened without changing the method of analysis, since at one 
point it will no longer be possible to analyze the rough surface as a 
one-dimensional interface. We believe that we are already past that limit at 
$R = 2$, but to give an estimate of the effects of the wall separation to 
the interested reader we have included the data for $R > 2$.

\section{SIMULATIONS\protect\\ }
\label{sec:level3} 
\begin{figure}[!]
\includegraphics[width=8cm]{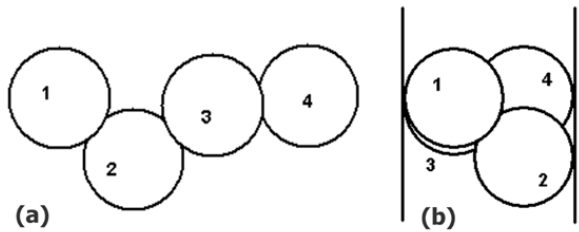}
\caption{A sample of the final surface obtained for $R = 1.6$. We have not 
shown the particles underneath the uppermost particles. (a) Front view, 
(b) side view of the deposited particles.} \label{fig3}
\end{figure}

We have also carried out computer simulations to investigate the effect of 
changing the cell width to the properties of rough surfaces formed by 
sedimentation. During the simulations we have deposited particles onto a 
quasi-two dimensional surfaces bounded by two walls. The separation between 
the walls (the width of the cell) is varied between $R = 1.1$ and $R = 1.9$. 
At the two ends of the cell we have used periodic boundary conditions. The 
particles are dropped onto the surface at random locations and once they 
touched the surface, they rolled to the local minimum which is reached when 
they are in contact with two other particles and a wall. A sample of the 
final surface is shown in (Fig.~\ref{fig3}). 

During the different runs we did not fix the length of the cell, but we 
varied the number of particles deposited with the requirement that the length 
of the final surface was about twice as large as the height. We have deposited 
100 surfaces of 5000, 10000, 50000, and 100000 particles each. We have 
observed that the final results did not significantly depend on the number 
of particles deposited. Figure 4 shows an example of the final surface 
deposited.

\begin{figure}[!]
\includegraphics[width=8cm]{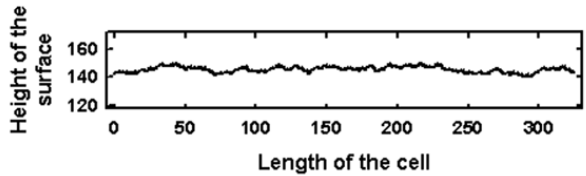}
\caption{The final surface after 50000 particles are deposited for the case 
$R = 1.1$. The length and the height of the surface are in units of $R$.} 
\label{fig4}
\end{figure}

\section{DISCUSSION\protect\\ }
\label{sec:level4}

As in the previous work, we have analyzed these rough surfaces using the 
scaling ansatz proposed by Family and Vicsek \cite{Family10790}. In this 
ansatz, the rms thickness of the interface is defined to be:

\begin{equation}
W (L,t) = \left[ \frac{1}{N} \sum_{i=1}^N \tilde{h}(x_{i},t)^{2}
\right]^ {\frac{1}{2}} \label{eqn1}
\end{equation}
where
\begin{equation}
   \tilde{h}(x_{i},t) = h(x_{i},t)-\bar{h}(t)
\label{eqn2}
\end{equation}
and
\begin{equation}
   \bar{h}(t) = \frac{1}{N} \sum_{i=1}^{N} h(x_{i},t) .
\label{eqn3}
\end{equation}

As discussed in the previous work, it is not at all clear that our system 
is in a scaling regime, nor is it obvious that scaling ideas should apply to 
sedimentation, but a useful way of analyzing our data is to adopt and extend 
the standard roughness analysis by tentatively accepting a scaling ansatz for 
rough interface growth. If we follow this ansatz, we expect that:

\begin{equation}
   W(L,t)= L^{\alpha} f(t/L^{\alpha / \beta})
\label{eqn4}
\end{equation}

where the exponents $\alpha$ and $\beta$ are the static and dynamic scaling 
exponents. The function $f(t/L^{\alpha / \beta})$ is expected to have an 
asymptotic form such that

\begin{equation}
   W(L,t) \sim t^{\beta} \; for \; t \ll L^{\alpha / \beta}
\label{eqn5}
\end{equation}
and
\begin{equation}
   W(L,t) \sim L^{\alpha} \; for \; t \gg L^{\alpha / \beta}
\label{eqn6}
\end{equation}

\begin{figure}[!]
\includegraphics[width=8cm]{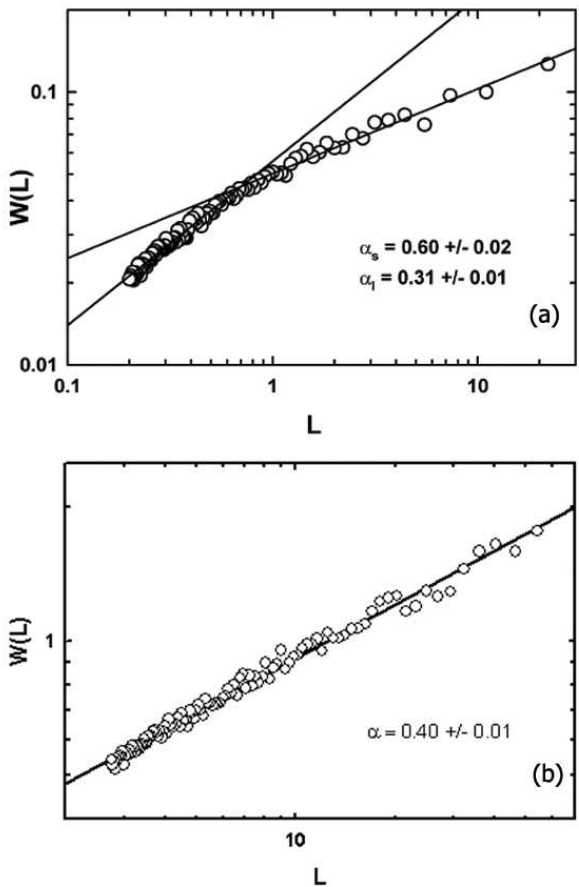}
\caption{Roughness function   versus L for a typical interface (a) at a 
gap of 0.8 mm in the experiment and (b) at $R = 1.6$ in the simulation.} 
\label{fig5}
\end{figure}

Fig.~\ref{fig5} shows an example of $W(L,t)$ at a typical gap/bead ratio 
in the experiments. To minimize the wall effects at the horizontal edges, 
we have used only the middle $70\%$ of each interface for our analysis.

\begin{figure}[!]
\includegraphics[width=8cm]{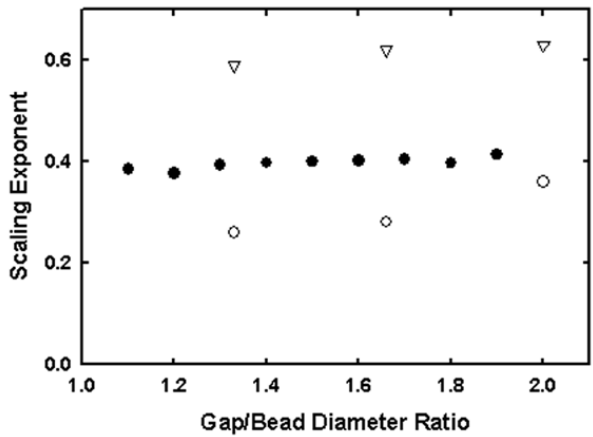}
\caption{The change of the average scaling exponents with gap/bead ratio, 
$R$. The empty circles ($\alpha_{s}$) and empty triangles ($\alpha_{s}$) 
denote the scaling exponents from the experiments. } 
\label{fig6}
\end{figure}

Again following the scaling ansatz, we find that at all values of $R$ 
(gap/bead ratio) studied, we still see two roughness exponents in the 
experiments, where $\alpha_{s}$ denotes the roughness exponent found at 
short length scales, while $\alpha_{l}$ denotes the roughness exponent found 
at long length scales. These roughness exponents have a crossover length scale 
at about 1 cm, which is typical from the previous work. Our earlier work 
corresponded to a gap width of 1.0 mm, and the experimental work gives 
similar results at this gap width as expected. As the value of $R$ is 
increased (Fig.~\ref{fig6}), we do not see any significant change in the 
value of either exponent. The simulation data on the other hand shows no 
crossover and yields one single roughness exponent. Just in the experiments 
we do not observe a change with changing gap/bead ratio, but the value of 
the scaling exponent obtained from the simulations is always  in between 
the two exponents obtained  from the experiments.

We must note that all of the previous discussion is based on the 
acceptance of the scaling arguments for this system. A careful review of 
the experimental data shows a crossover, but the two scaling regimes are 
not clearly linear (Fig.~\ref{fig5}(a)). An alternate argument can be made 
that this data does not show clear evidence of scaling. If there is scaling, 
there are two exponents, but the cross-over between the two length scales 
does not appear to be sharply defined. Although it is clear that the behavior 
at small length scales seems different than that at large length scales, an 
argument can certainly be made that the data change continuously at different 
length scales. Therefore the data in the simulations show a clearer picture 
as being the mathematical equivalent of the experimental problem. It has already 
been noted that even in the absence of particle-particle interactions, 
particle-wall interactions can play a significant role in determining the final 
structure of the surface. Also we have ignored effects coming from weight of the 
particles and also their kinetic energy. In the simulations the particles stopped 
when they came into contact with two other particles and a wall whereas in the 
experiments we have observed that some particles did not stay at the local minima 
and continued towards a global minimum with the effect of their momentum. Thus 
the simulations should be regarded as a limiting case of infinite viscosity and 
zero downward momentum.

We have investigated the effect of the interaction between the walls of the 
container and the sedimenting particles on the roughness exponent of the surface 
formed by this quasi-two-dimensional sedimentation. If the scaling ansatz is 
accepted, the roughness exponent is found to be robust to the changes in the 
separation between the walls of the container as observed in the experiments 
and by simulating the same problem. We have been unable to reproduce the slight 
increase in the roughness exponent at $R = 2$ using computational methods. As the 
experimental data show evidence of continuous change with the lengthscale $L$, 
the possibility that scaling arguments do not hold should be taken seriously. In 
contrast, the simulation data show good agreement with the scaling arguments 
suggesting that particle-wall interactions can be blamed for the deviation from 
the scaling behavior.

\begin{acknowledgements}
This project was funded by grants from the Research Foundation, the Office of 
Naval Research, and the Department of Defense / DTRA Environmental Management - 
Bioenvironmental Hazards Research Program.   
\end{acknowledgements}

\end{document}